\documentclass[10pt,letterpaper,twocolumn]{article} %% two column, final layout

\usepackage{ol2}
\usepackage{amsmath,amssymb,graphicx}
\usepackage{hyperref}
\usepackage{enumitem}
\usepackage{color}

\newcommand{\fref}[1]{Fig.~\ref{fig:#1}}
\newcommand{\eref}[1]{(\ref{eq:#1})}

\newcommand{\sub}[1]{_{\mathrm{#1}}}
\newcommand{\super}[1]{^{\mathrm{#1}}}

\begin{document}
\twocolumn[ %% activate for two-column option

%\title{Optical resonator sensing and control signals derived from
%  transverse spatial modes of order two}

\title{Length control of an optical resonator using second-order
  transverse modes}

%\title{A simple method of deriving optical cavity sensing and control
%   signals based on transverse spatial modes of order two}

\author{John Miller$^*$ and Matthew Evans}
\address{Massachusetts
   Institute of Technology, Cambridge, MA 02139,
   USA\\$^*$Corresponding author: jmiller@ligo.mit.edu}

\begin{abstract}
  We present the analysis of an unorthodox technique for locking a
  laser to a resonant optical cavity. Error signals are derived from
  the interference between the fundamental cavity mode and
  higher-order spatial modes of order two excited by mode
  mismatch. This scheme is simple, inexpensive and, in contrast to
  similar techniques, first-order-insensitive to beam jitter. After
  mitigating sources of technical noise, performance is fundamentally
  limited by quantum shot-noise.
\end{abstract}

% In the absence of other sources of noise, shot-noise-limited
% performance is achievable.

\ocis{120.2230, 140.4780, 120.3180.}
% Fabry-Perot (Instrumentation, measurement, and metrology), optical
% resonators (lasers and laser optics) and interferometry
% (Instrumentation, measurement, and metrology)

] %% activate for two-column option

%\section{Introduction}
Due to their power gain and non-linear phase response, resonant
optical cavities are routinely employed across a variety of
fields. However, to exploit these useful properties, cavities must be
held resonant with their input laser fields using a feedback control
system, a process known as \emph{locking}.

In order to achieve lock, a suitable error signal, describing the
offset of the incoming laser light from resonance, must be
generated. This procedure ordinarily involves comparing light which
interacts strongly with the cavity to a stable reference. Common
references include a fixed voltage source (side-of-fringe locking) and
audio- or radio-frequency modulation sidebands (dither and
Pound-Drever-Hall locking respectively) \cite{Boyd1996}. More
infrequently, polarisation effects are also used to create locking
references (see e.g.~\cite{Hansch1980,Asenbaum2011}).

This Letter analyses the use of a higher-order Laguerre-Gaussian
($LG$) mode of order two \cite{Siegman1986}, excited by mode mismatch,
as an alternative reference for cavity locking. We introduce a
theoretical scheme for generating error signals from this reference
and discuss two practical methods of implementing it. Experimental
investigation confirms our predictions.

% As recognised previously \cite{Wieman1982}, techniques based on
% higher-order modes (HOMs) are particularly appealing because they
% remove the need for external references or modulation-demodulation
% stages, do not require small-aperture modulators in the beam path and
% allow one to lock to the top of a resonance peak. Hence, these
% techniques are relatively simple and inexpensive, consume little
% electrical power, are ideal for high-power applications and offer good
% noise performance.
As recognised previously \cite{Wieman1982}, techniques based on
higher-order modes (HOMs) are particularly appealing because they
remove the need for external references or modulation-demodulation
stages whilst maintaining the ability to lock to the top of a resonance
peak. Hence these techniques offer good sensitivity, are relatively
simple and inexpensive, consume little electrical power, are
low-weight, do not require small-aperture modulators in the beam path
and are robust against temperature fluctuations. In addition to
general laboratory applications, these properties make HOM-based
techniques ideal for field-deployable (including satellite-based)
instruments, high-power systems and experiments requiring multiple
control loops.

\begin{figure}[thbp!]
  \centering
  \includegraphics[width=\columnwidth]{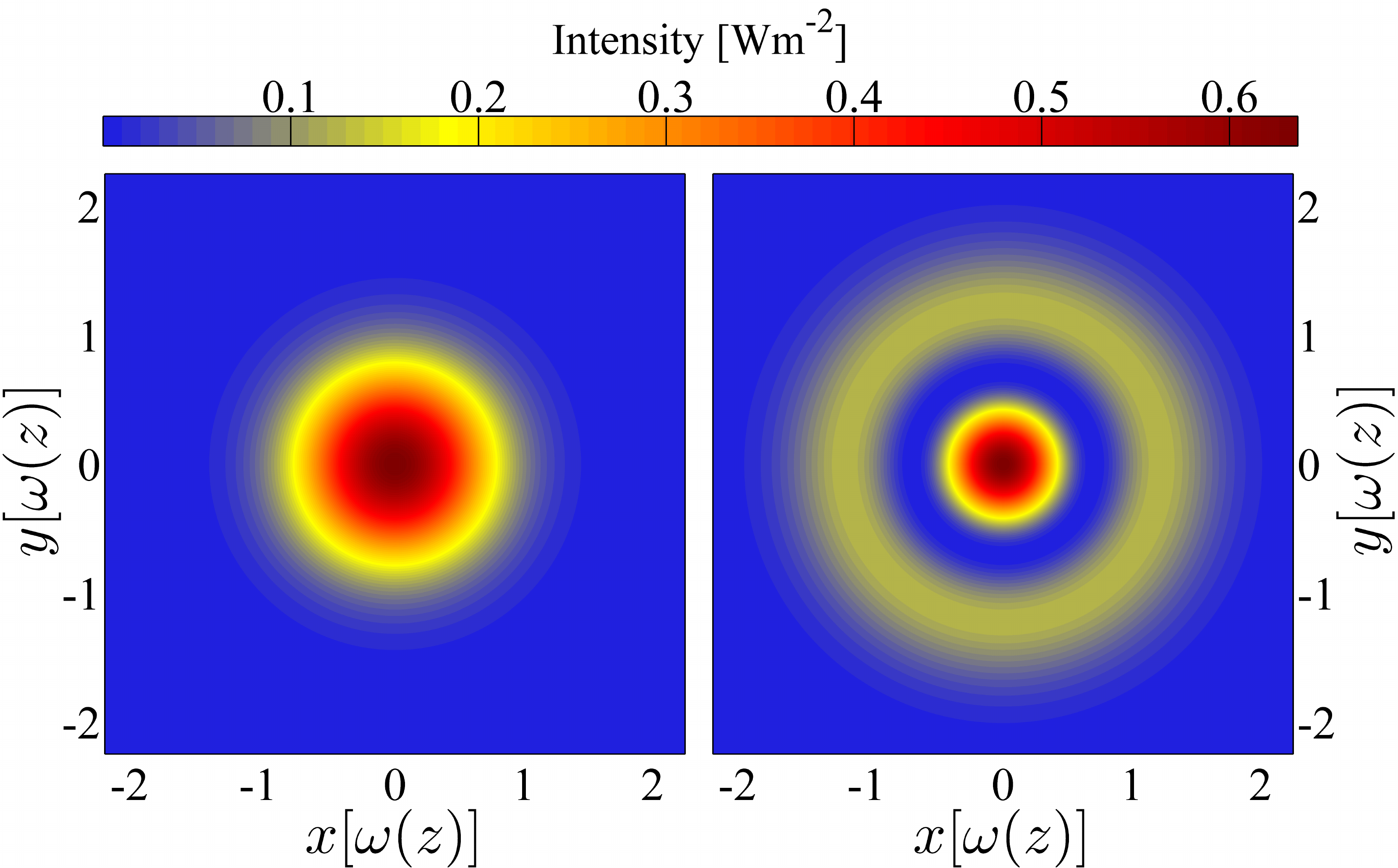}
  \caption{Optical intensities of the $\Psi_0$ and $\Psi_2$ spatial
    modes (left and right axes respectively). Each mode has an
    integrated power of 1 W.}
  \label{fig:modes}
\end{figure}
Moreover, in contrast to prior investigations \cite{Shaddock99}, which
made use of an odd-order HOM as a locking reference, our technique is
first-order insensitive to beam spot motion at the detector. This
robustness is particularly beneficial in industrial and
suspended-mirror environments.

%\section{Mode mismatch and higher-order modes}
An optical cavity will decompose any input beam into its cavity
eigenmodes. Our analysis considers a rotationally-symmetric input beam
which is perfectly aligned to the axis of a spherical-mirror cavity
(the $z$ axis) but whose beam parameters differ slightly from those of
the fundamental cavity mode. In this case, the cavity eigenmodes
excited by the mismatched input beam are well-approximated by the
family of $LG_{m,n=0}$ modes.

\begin{figure}[thbp!]
  \centering
  \includegraphics[width=\columnwidth]{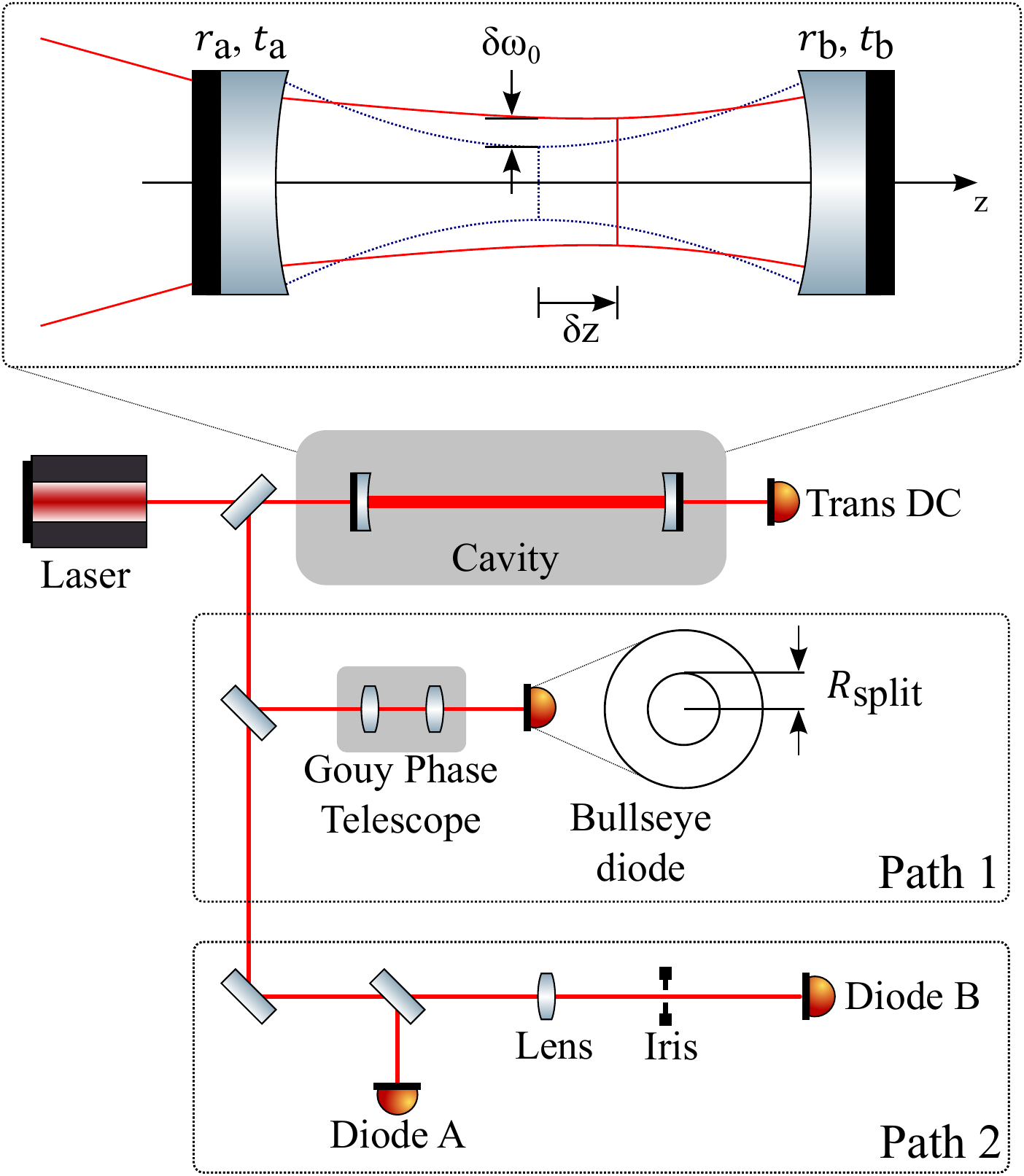}
  \caption{The optical setup required to implement our technique. Only
    one of the two paths in reflection of the cavity, either 1 or 2,
    is required. Path 1's bullseye photodiode consists of an inner circle
    of radius $R\sub{split}$ surrounded by an annulus. The expanded
    view of the cavity depicts the fundamental cavity eigenmode using
    a dashed blue line and the mismatched input beam in red. Vertical
    lines indicate the waist positions and diameters.}
  \label{fig:apparatus}
\end{figure}
For small (first order in the perturbation) mode mismatches only
coupling to the two lowest order modes need be considered. Neglecting
common phase factors irrelevant to our analysis, these modes may be
expressed as
\begin{align}
\Psi_0(r,z)&=\sqrt{2/\pi\omega^2(z)}\exp[-r^2/\omega^2(z)],\\
\Psi_2(r,z)&= \sqrt{2/\pi\omega^2(z)}[1-2r^2/\omega^2(z)]\nonumber\\
&\quad\times\exp[-r^2/\omega^2(z)+i2\gamma(z)]\nonumber\\
&=\widehat{\Psi}_2\exp[i2\gamma(z)],
\end{align}
where $\omega(z)=\omega_0[1+(z/z\sub{R})^2]^{1/2}$ is the beam spot
size ($\omega_0$ being the beam waist;
$z\sub{R}=\pi\omega_0^2/\lambda\sub{laser}$, the Rayleigh range),
$\gamma(z)=\mathrm{tan}^{-1}(z/z_R)$ is the Gouy phase and subscripts
indicate the mode order $2m+n$. The intensity distribution of each of
these modes is shown in \fref{modes}.

Consider a cavity with waist $\omega\super{cavity}_0$ at position
$z=0$. In the basis of cavity eigenmodes, a 1 W, fundamental mode,
input beam with waist size $\omega_0\super{in}$ at position $z=\delta
z$ (see \fref{apparatus}) may be described as
\begin{align}
  \label{eq:expansion}
    \Psi_0\super{in}&=\Psi_0\super{cavity}+ \epsilon \Psi_2\super{cavity},\\
\mathrm{where}\quad\quad
 \epsilon &= \frac{\delta \omega_0}{\omega\super{cavity}_0}+ i \frac{\delta z}{2 z_R\super{cavity}},
\end{align}
and $ \delta\omega_0 = \omega\super{in}_0-\omega\super{cavity}_0$
\cite{Anderson84}.
% \begin{equation}
%   \label{eq:expansion}
%     \Psi_0\super{in}=\Psi_0\super{cavity}+ \epsilon \Psi_2\super{cavity},
% \end{equation}
% where
% \begin{equation}
%  \epsilon = \frac{\delta \omega_0}{\omega\super{cavity}_0}+ i \frac{\delta z}{2 z_R\super{cavity}},
% \end{equation}
% and $ \delta\omega_0 = \omega\super{in}_0-\omega\super{cavity}_0$
% \cite{Anderson84}.

%\section{Cavity response and error signal}
%\label{sec:responseAndError}
Our scheme generates error signals via the light reflected from a
resonant optical cavity. The complex amplitude reflectivity of a
cavity is given by
\begin{equation}
r(\phi\sub{rt})=r\sub{a}-\frac{t\sub{a}^2r\sub{b}\exp(-i\phi\sub{rt})}{1-r\sub{a}r\sub{b}\exp(-i\phi\sub{rt})},
\end{equation}
where $\phi\sub{rt}$ is the phase acquired upon one cavity round-trip,
$r_a$ and $t_a$ are the amplitude reflection and transmission
coefficients of the input mirror and $r_b$ is the reflection
coefficient of the end mirror.

In stable cavities the eigenmodes are not degenerate and each
experiences a different cavity response by virtue of the Gouy phase
contribution to $\phi\sub{rt}$.
% \begin{equation*}
%   (2m+n+1)\arccos[\mathrm{sgn}(g_1)\sqrt{g_1g_2}].
% \end{equation*}
Close to a fundamental-mode resonance, the second-order mode interacts
very weakly with the cavity and hence is available as a stable
reference to which the fundamental mode can be compared. This effect
has previously been used to create automatic alignment and
mode-matching systems \cite{Anderson84, Morrison94A, Mueller00}.

Applying the cavity response to \eref{expansion} and detecting the
reflected power on a photodiode of area $\mathcal{A}$ we have, in
  the cavity frame,
\begin{equation}
  \label{eq:Prefl}
    P\sub{refl}\super{cavity}=\int_\mathcal{A} \underbrace{|r_0\Psi_0|^2}\sub{(i)}+\underbrace{|\epsilon r_2\Psi_2|^2}\sub{(ii)}+\underbrace{2\Re(\epsilon^*r_0r_2^*\Psi_0\Psi_2^*)}\sub{(iii)} \, \mathrm{d}\mathcal{A}
\end{equation}
where $r_i$ is the cavity reflectivity for $\Psi_i$ (here and
henceforth we omit explicit dependence on $\phi\sub{rt}$) and $\alpha^*$
and $\Re(\alpha)$ denote the complex conjugate and real part of $\alpha
\in\mathbb{C}$ respectively.

The reflected signal comprises three components. Terms (i) and (ii)
yield absorption-like features around the $\Psi_0$ and $\Psi_2$
resonant frequencies, respectively, and are constant elsewhere. Term
(ii) may be ignored with impunity so long as the $\Psi_0$ and $\Psi_2$
resonances do not overlap. The term of interest, describing the
interference between our reference $\Psi_2$ mode and the fundamental
cavity mode $\Psi_0$ is (iii). Converting all complex quantities into
polar coordinates this term may, more instructively, be written
\begin{equation}
 \label{eq:error}
2\,\mathcal{I}|\epsilon
  r_0r_2|\sin[\theta_{r_2}-\theta_{r_0}+2\gamma(z) +\theta_\epsilon+\pi/2]
\end{equation}
where $\mathcal{I}=\int_\mathcal{A}
\Psi_0\widehat{\Psi}_2\,\mathrm{d}\mathcal{A}$ and $\theta_{\alpha}$
denotes $\arg(\alpha)$.
%argument of $\alpha$.

As desired, this signal is sensitive to the difference between the
phase of the reflected fundamental mode and that of the second-order
mode, $\theta_{r_2}-\theta_{r_0}$. On passing through a
fundamental-mode resonance, $\theta_{r_0}$ changes rapidly whereas
$\theta_{r_2}$ is essentially fixed and zero, therefore acting as a
stable reference.

%Gouy phase at the detector
$\theta_\epsilon$ describes the `flavour' of the cavity mismatch. For
pure waist position (size) mismatches this quantity is $\pm\pi/2$ (0
or $\pi$). By controlling the Gouy phase, $\gamma(z)$, either through
simple propagation or the construction of an appropriate telescope,
the influence of this and the remaining terms may be removed to yield
a quantity proportional to $\sin(\theta_{r_0})$ -- a bi-polar
function, centred about cavity resonance, which is ideal for use as an
error signal in a feedback control system.

%\section{Photodiode geometry}
%\subsection{Conventional diode}
In order to isolate the error signal, care must be taken over the
choice of photodiode geometry. Since we are examining interference between
orthogonal modes, detection over a conventional single-element photodiode
will yield no signal, i.e.~$\mathcal{I}=\int^\infty_0  \Psi_0\widehat{\Psi}_2\,2\pi r\,\mathrm{d}r =0$.
% \begin{equation}
%   \label{eq:integral}
% \mathcal{I}=\int\limits^\infty_0  \Psi_0\widehat{\Psi}_2\,2\pi r\,\mathrm{d}r =0.
% \end{equation}
%This result can be understood through closer examination of the
%integrand (see \fref{integrand}).

%\subsection{Bullseye diode}
The orthogonality of the interfering modes may be circumvented by any
photodiode which does not sample all of the incident light. However,
following the distribution of the electric field, a logical choice is
to use a two-part, radially-split, \emph{bullseye} photodiode (see
\fref{apparatus}, Path 1) and construct signals proportional to
\begin{equation}
  \label{eq:integration}
  % \mathcal{I}=\int\limits^{R\sub{split}}_0
  % \Psi_0\widehat{\Psi}_2\,2\pi r\,\mathrm{d}r -
  % \int\limits^\infty_{R\sub{split}}  \Psi_0\widehat{\Psi}_2\,2\pi
  % r\,\mathrm{d}r
  \mathcal{I}=\int^{R\sub{split}}_0  \Psi_0\widehat{\Psi}_2\,2\pi r\,\mathrm{d}r -  \int^\infty_{R\sub{split}}  \Psi_0\widehat{\Psi}_2\,2\pi r\,\mathrm{d}r
\end{equation}
by subtracting the outputs of the two segments.

Specifically, the error signal is maximised by taking
$R\sub{split}=\omega(z)/\sqrt{2}$, as this is the point where the
electric field of $\widehat{\Psi}_2$ changes sign, hence obviating any
%cancellation of signal (see \fref{integrand}). With this choice
%$\mathcal{I}=2/e$.
cancellation of signal. With this choice $\mathcal{I}=2/e$.

However, since the $\Psi_0$ power on the two segments of the photodiode is
not balanced, operating in this way can return impure error signals
due to contamination by term (i) of \eref{Prefl}. Hence, this mode of operation is
best reserved for strongly overcoupled, high-finesse cavities.
%Small dip

%Moreover, any laser power fluctuations will also affect performance.

% The error signal is offset. Either lock where error signal=0 and
% get detuning but no coupling to laser power noise OR lock at the
% peak of resonance and have a coupling.

%contamination by term (i) % effect
For more general cavity configurations, this effect is easily
mitigated by setting $R\sub{split}=\omega(z)\sqrt{\log(2)/2}$ to
balance the $\Psi_0$ power on each photodiode segment. In this case
$\mathcal{I}=\log(2)$, representing only a $\sim$6\% loss of
signal.

Removing the influence of variations in $\Psi_0$ power necessarily
dictates that changes in $\Psi_2$ power are discernible. In
particular, the tail of the $\Psi_2$ resonance can cause small error
signal offsets at the $\Psi_0$ lock point. Again, such offsets are
negligible so long as the $\Psi_0$ and $\Psi_2$ resonances do not
overlap significantly. For example, using the parameters given in
\fref{errorSignals}, this effect produces an offset which reduces
circulating power by $\sim$5~ppm.

%\subsection{Two conventional diodes}
Although bullseye photodiodes are available they are by no means
commonplace. Hence, we now introduce an alternative means of
implementing our scheme (see \fref{apparatus} Path 2). Error signals
obtained using this implementation are shown in \fref{errorSignals}.

The light reflected from the optical cavity is divided at a
beamsplitter and relayed to two standard single-element photodiodes. Photodiode
A captures the entire cross-section of the reflected beam, and is
therefore sensitive to term (i), allowing us to subtract the
contribution of the $\Psi_0$ resonance dip from our error signal.

The beam propagating towards Photodiode B is clipped significantly by an
adjustable iris. Thus, Photodiode B plays the role of the central part of a
bullseye photodiode and is sensitive to term (iii).

\begin{figure}[htbp!]
  \centering
  \includegraphics[width=\columnwidth]{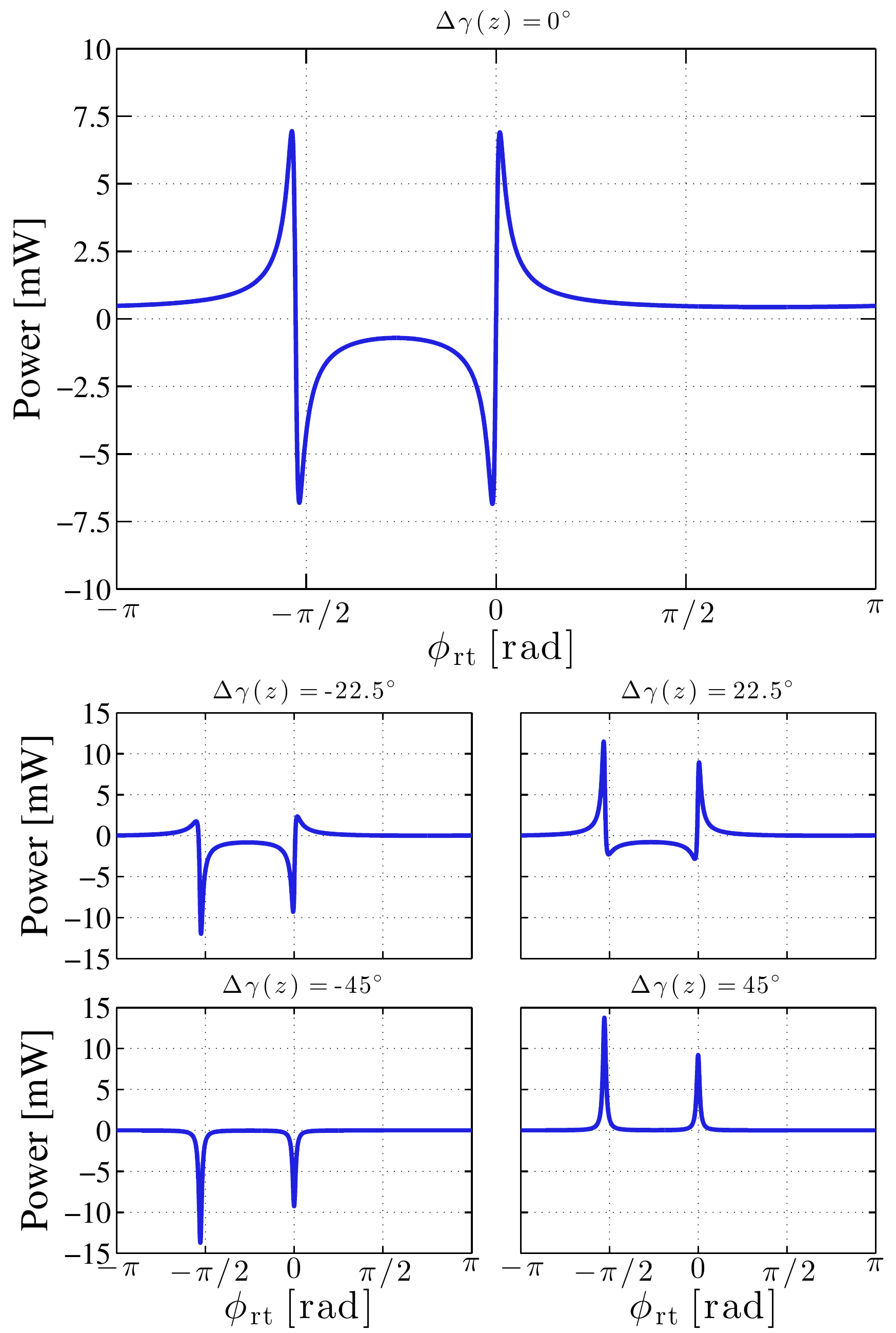}
  \caption{Uppermost axes -- Theoretical error signal obtained using
    two conventional photodiodes and the following cavity parameters;
    $P\sub{laser}=1$~W, $\lambda\sub{laser} = 1064$~nm, $L\sub{cavity}
    = 1.3$~m, $r_a^2 = 0.95$, $r_b^2 = 0.99$, ROC$_a$ = ROC$_b$ = 4~m,
    $\epsilon = -0.01+i0.005$. The width of the linear part of the
    error signal is set by the full-width-half-maximum-power cavity
    linewidth. The feature near to $\phi\sub{rt}=-\pi/2$ is an error
    signal for the $\Psi_2$ resonance. In this case the fundamental
    mode acts as the phase reference. Lower axes -- Error signals
    during tuning of iris position for various deviations,
    $\Delta\gamma(z)$, from the optimal Gouy phase. At each location
    the radius of the iris is adjusted such that the error signal is
    zero far from resonance.}
  \label{fig:errorSignals}
\end{figure}
The final error signal for use in a feedback control or measurement
system is of the form \hbox{$\mbox{Photodiode B}-c\times\mbox{Photodiode A}$},
where the constant $c\in\mathbb{R}$ is chosen to null sensitivity to
term (i) and depends on photodiode responsivities and transimpedance
gains. For matched photodiodes, $c=1-\exp[-2R\sub{iris}^2/\omega^2(z)]$.

\begin{figure}[htbp]
  \centering
  \includegraphics[width=0.8\columnwidth]{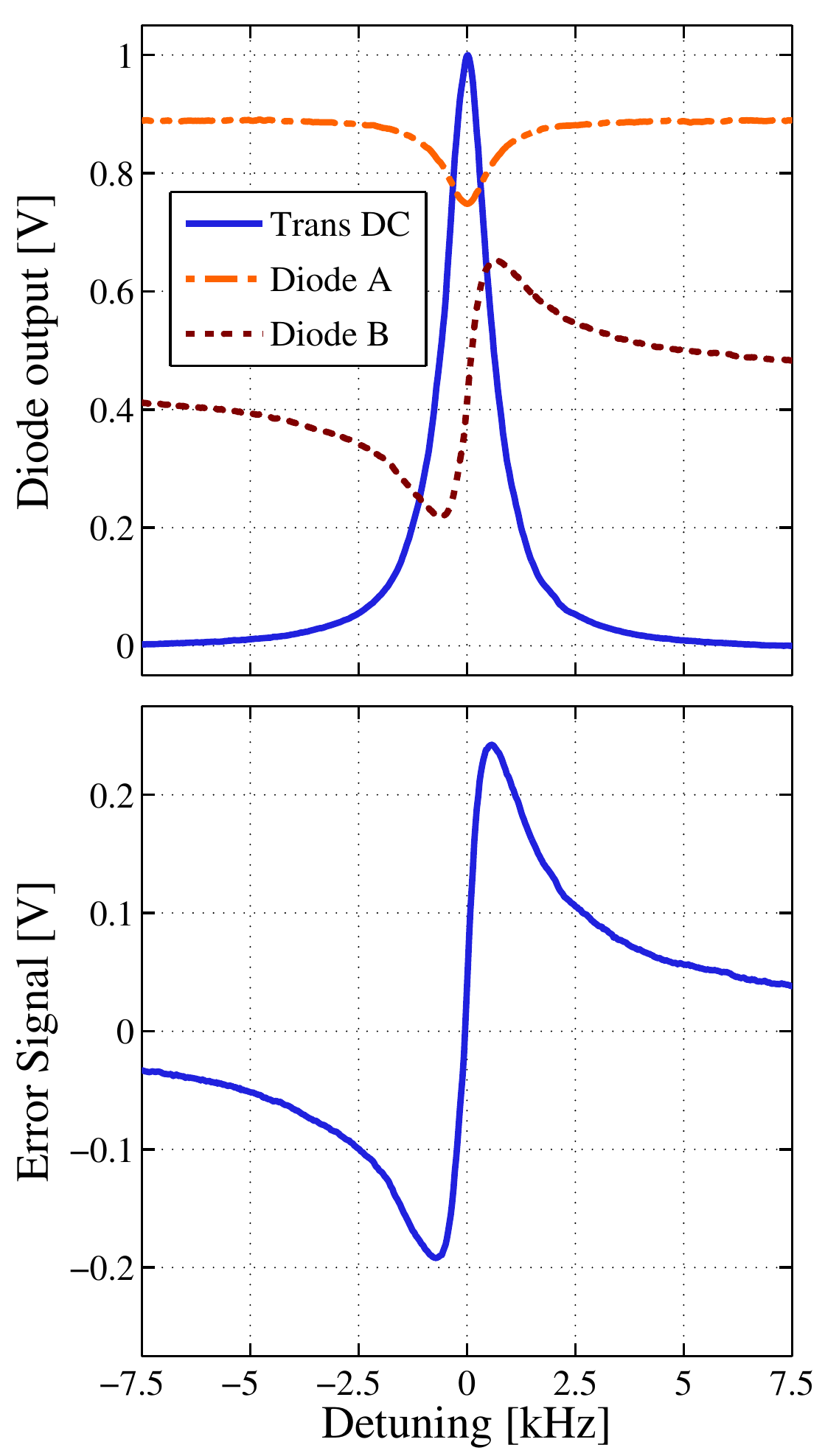}
  \caption{Output of three photodiodes shown in \fref{apparatus}
    (upper axes) and the resulting error signal (lower axes) as the
    laser frequency is swept across $\Psi_0$ resonance. Trans DC has
    been normalised and Photodiode B's response has been scaled to
    match that of Photodiode A. At the time of this measurement, 70\%
    of the input light was coupled into the $\Psi_0$ mode. Such large
    mode mismatch is not a requirement of this technique.}
  \label{fig:experimentalSignals}
\end{figure}
The Gouy phase and spot size at a bullseye detector are
interdependent. The use of an adjustable iris as the effective
clipping aperture eliminates this difficulty, with the Gouy phase at
the iris now defining the relative phase of the interfering spatial
modes. This phase is easily controlled by placing a waist-forming lens
upstream of the aperture such that the beam explores a wide range of
Gouy phases over a reasonable propagation distance. Modifying the
position and radius of the aperture independently, one quickly
approaches the optimal configuration given by, assuming
\hbox{$\theta_{r_2}=0$},
\hbox{$\gamma(z\sub{iris})=k\pi/2-\pi/4-\theta_\epsilon/2,\,
  k\in\mathbb{Z}$} and, as before,
$R\sub{iris}=\omega(z)/\sqrt{2}$. The lower axes of
\fref{errorSignals} illustrate the signals one might observe during
this tuning process. The accuracy with which the optimal configuration
can be achieved in practice could require further examination if this
technique were to be used in exacting applications such as
metrological standards.

Using two conventional photodiodes removes the need for specialised
hardware at the expense of sensitivity. Since we do not detect all of
the light, our signal is reduced by 50\% with respect to optimal
detection on a bullseye photodiode.

%\section{Experimental demonstration}
To validate our theoretical analysis, we applied the two-photodiode
version of our scheme to a high-finesse optical cavity. The technique
was found to be extremely robust against errors in iris radius and
placement. Experimental data are shown in
\fref{experimentalSignals}. Future practical investigations should
explore the long-term stability of this method.

%\section{Discussion and conclusions}
%\subsection{Comparison with tilt locking}
A comparable technique, relying on an $HG_{10}$ mode
\cite{Siegman1986} as the phase reference, has been developed
previously \cite{Shaddock99}. Although this alternative offers an 8\%
%sqrt(2/pi) against 2/e
improvement in theoretical sensitivity, we believe that the scheme
described herein is superior for a number of reasons.

Foremost amongst these its insensitivity to beam jitter in reflection
of the cavity. Odd-order-mode techniques are linearly sensitive to
beam motion. This dependency can be mitigated but not without
increasing the complexity of the scheme (`double-pass tilt locking'),
somewhat limiting its appeal. In contrast the sensitivity for
even-order-mode techniques, such as ours, is quadratic. To appreciate
this consider that, for the circularly symmetric modes under
discussion, the power transmitted through an iris decreases with
relative beam-iris motion, independent of the direction of the motion.

Our scheme does introduce a new sensitivity to beam spot size
changes. However, such changes are rare and occur on thermal
timescales whereas alignment fluctuations are common and exist over a
wide range of frequencies.

Odd modes are excited by cavity misalignment while even modes arise
due to mode mismatch. In general, achieving excellent cavity alignment
is undemanding while attaining equivalent mode-matching efficiency
requires extraordinary measures \cite{Mueller00}. Therefore,
pre-existing second-order modes suitable for use as a locking
reference are almost invariably present, even after a system has been
finely adjusted, and do not need to be purposely introduced. In
contrast, exciting an additional $HG_{10}$ reference mode requires
that the cavity be intentionally misaligned, establishing noise
coupling pathways and reducing shot-noise limited performance.
%static misalignment couples input beam jitter to 
%optimised
%thoroughly tuned

% Odd modes are excited by cavity misalignment while even modes arise
% due to mode mismatch. In general, achieving excellent cavity alignment
% is undemanding while attaining equivalent mode-matching efficiency
% requires extraordinary measures \cite{Mueller00}. Therefore, exciting
% a reference $HG_{10}$ mode requires that the cavity be purposely
% misaligned, introducing noise coupling pathways and reducing
% shot-noise limited performance. In contrast, even order modes are
% invariably present even after a system has been finely adjusted.
% %static misalignment couples input beam jitter to 
% %optimised
% %thoroughly tuned

%\subsection{Conclusions}
We have presented the analysis of an alternative means of locking a
laser to a resonant optical cavity. This technique offers the
possibility of shot-noise-limited performance whilst remaining
uncomplicated and low-cost. Experimental investigation has confirmed
our theoretical calculations and shown the technique to be simple to
apply and robust against errors in implementation.

%\section*{Acknowledgements}
The authors gratefully acknowledge the support of the National Science
Foundation and the LIGO Laboratory, operating under cooperative
Agreement No. PHY-0757058. They also acknowledge Tomoki Isogai,
Patrick Kwee and Lisa Barsotti for their roles in developing the
apparatus on which the experimental portion of this work was performed.
This paper has been assigned LIGO Document No. LIGO-LIGO-P1300209.

%\bibliographystyle{ol} % Bibliography style file
%\bibliography{/Users/miller/Documents/Misc/Bibliographies/MillerBibliography} % Bibliography database file

\pagebreak

\section*{Informational Fifth Page}
% In this section, please provide full versions of citations to
% assist reviewers and editors (OL publishes a short form of
% citations) or any other information that would aid the peer-review
% process.

%\bibliographystyle{osajnl} % Bibliography style file
%\bibliography{/Users/miller/Documents/Misc/Bibliographies/MillerBibliography} % Bibliography database file

\end{document}